\documentstyle[11pt,newpasp,twoside,epsf]{article}
\def\edcomment#1{\iffalse\marginpar{\raggedright\sl#1\/}\else\relax\fi}
\reversemarginpar

\begin{document}
\title{Stellar Exotica produced from Stellar Encounters}
 \author{Melvyn B. Davies}
\affil{Department of Physics \& Astronomy, University of Leicester, Leicester
LE1 7RH, UK.}

\begin{abstract}
The importance of stellar encounters in producing stellar
exotica in dense stellar
clusters is reviewed.
We discuss how collisions between main--sequence stars may be 
responsible for the 
production of blue stragglers in
globular clusters. We also discuss the possible pathways
to the production of X--ray binaries, cataclysmic variables, and millisecond
pulsars in globular clusters.
Neutron stars in globular clusters
are likely to exchange into binaries containing moderate--mass
main--sequence stars, replacing the lower--mass components of the original
systems. These binaries will become intermediate--mass X--ray binaries
(IMXBs), once the moderate--mass star evolves off the main--sequence, as
mass is transferred onto the neutron star possibly spinning it up in the
process. Such systems may be responsible for the population of
millisecond pulsars
(MSPs) that has been observed in globular clusters. Additionally,
the period of mass--transfer (and thus X--ray visibility) in the vast
majority of such systems will have occurred 5 -- 10 Gyr ago
thus explaining the  observed relative paucity of X--ray binaries today,
given the large MSP population.
\end{abstract}

\section{Introduction}

Stellar clusters are ubiquitous. 
Globular clusters contain some of the
oldest stars, whilst the youngest stars are found in OB associations
or in other clusters associated with recent star formation.
Such crowded places are hostile environments: a large fraction
of stars will collide or undergo close encounters. Wide binaries
are likely to be broken up, whilst tighter ones will suffer major
pertubations and possibly collisions from passing stars.
Hydrodynamical computer simulations of such encounters
play a vital role 
in understanding how collisions will affect the evolution of stellar
clusters and produce the myriad of stellar exotica seen such as
X--ray binaries, blue stragglers and millisecond pulsars.
The cluster of stars at the centre of a galaxy may provide the
material to form a massive black hole and fuel it as a quasar.
Encounters in very young clusters will influence
the fraction of stars contained in binaries and their nature.
Such encounters will affect the stellar population of the entire
galaxy as all stars are formed in clusters.

In this review, we consider the importance  of stellar
encounters within globular clusters, where collisions
between two main--sequence stars may produce blue stragglers.
Encounters involving neutron stars and non--compact stars
may help explain the large population of millisecond pulsars.
In $\S$2 we compute the timescales for encounters
in dense stellar clusters. In $\S$3 the importance of such encounters
is discussed. 
Encounters in globular clusters are considered
in $\S\S$4, 5, and 6, where we discuss blue stragglers,
low--mass X--ray binaries, and cataclysmic variables,
respectively. A summary of this review is given in $\S$7.

\begin{figure}[ht]
\begin{center}
\plotfiddle{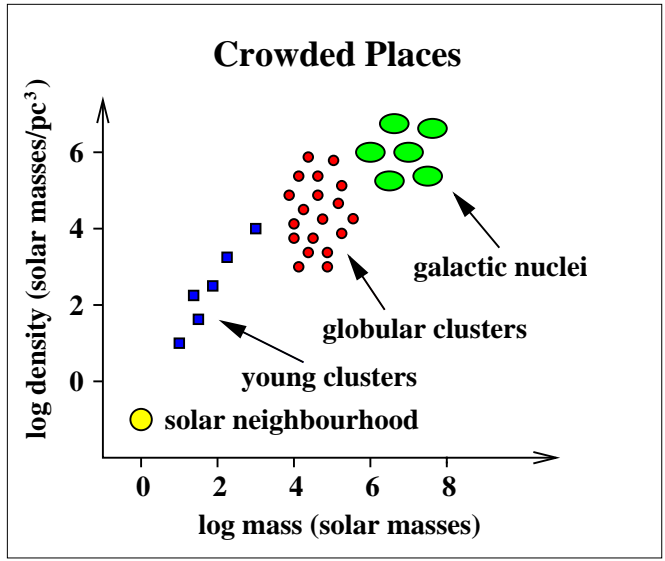}{8truecm}{0}{40}{40}{-130}{0}
\caption{\label{mdfig1} A schematic diagram of crowded places}
\end{center}
\end{figure}

\section{Encounter Timescales}

Encounters between two stars will be extremely rare in the low--density
environment of the solar neighbourhood. However, in the cores of globular
clusters, and galactic nuclei, number densities are sufficiently high 
($\sim 10^5$ stars/pc$^3$ in some systems, as shown in Figure 1)
 that
encounter timescales can be comparable to, or even less than, the 
age of the universe. In other words, a large fraction of the stars
in these systems will have suffered from at least one close encounter
or collision in their lifetime.

The cross section for two stars, having a relative velocity at infinity
of $V_\infty$, to pass within a distance $R_{\rm min}$ is given by

\begin{equation}
\sigma = \pi R_{\rm min}^2 \left( 1 + {V^2 \over V_\infty^2} \right)
\end{equation}

\noindent where $V$ is the relative velocity of the two 
stars at closest approach in a parabolic encounter ({\it i.e.\ } 
$V^2 = 2 G (M_1 + M_2)/R_{\rm min}$, where $M_1$ and $M_2$ are the masses
of the two stars).
The second term is due to the attractive gravitational force, and
is referred to as gravitational focussing.
In the regime where $V \ll V_\infty$ (as might be the case in 
galactic nuclei with extremely high velocity dispersions), we recover
the result, $\sigma \propto R_{\rm min}^2$. However, if $V \gg V_\infty$
as will be the case in systems with low velocity dispersions, such as
globular clusters, $\sigma \propto R_{\rm min}$.

One may estimate the timescale for a given star to undergo an encounter
with another star,
$\tau_{\rm coll} = 1/n \sigma v$. For clusters with low velocity dispersions,
we thus obtain

\begin{equation}
\tau_{\rm coll} = 7 \times 10^{10} {\rm yr} \left( {10^5pc^{-3} \over
n } \right) \left( { v_\infty \over 10km/s } \right) 
\left( { R_\odot \over R_{\rm min} } \right) \left( { M_\odot \over
M } \right) \ {\rm for} \ v \gg v_\infty 
\end{equation} 

\noindent where $n$ is the number density of single stars of mass $M$. 
For an encounter between two single stars to be hydrodynamically
interesting, we typically require $R_{\rm min} \sim 3 R_\star$ 
(see for example, Davies, Benz \& Hills 1991). 
We thus see that for typical globular clusters, where $n \sim 10^5$,
up to 10\% of the stars in the cluster cores will have undergone a
collision at some point during the lifetime of the cluster.

We may estimate the timescale for an encounter between a binary and a third,
single star, in a similar manner where now $R_{\rm min} \simeq d$, 
where $d$ is the semi--major axis of the binary. The encounter
timescale for a binary may therefore be relatively short as the semi--major
axis can greatly exceed the size of the stars it contains.
For example, a binary with $d \sim$ 1AU (ie 216
R$_\odot$), will have an encounter timescale $\tau_{\rm enc} 
\ll 10^{10}$ years in the core of a dense globular cluster. 
Thus encounters between binaries and single stars may be important
in stellar clusters even if the binary fraction is small.

Encounters between single stars and extremely wide binaries 
will lead to the break up of the binaries as the kinetic energy of 
the incoming star exceeds the binding energy of the binary. Such
binaries are often referred to as being {\sl soft}.
Conversely, {\sl hard} binaries will be resilient to break up. 
Encounters between single stars and hard binaries have three main outcomes
as shown in Figure 2.
A fly--by may occur where the incoming third star leaves the original
binary intact. However such encounters will harden (ie shrink) the binary, and 
alter its eccentricity. Alternatively, an exchange may occur where the 
incoming star replaces one of the original components of the binary.
During the encounter, two of the stars may pass so close to each other
that they merge or form a very tight binary (as they raise tides in each
other). The third star may remain bound to the other two as indicated in
outcomes d), e) and f) in Figure 2.

The cross section for a single star to pass within a distance $R_{\rm min}$
of the center of mass of a binary is given
by $\sigma = \pi R_{\rm min}^2 (1 + V_c^2/V_\infty^2)$, where $V_\infty$ is the
relative speed at infinity, and $V_c$ is the relative speed at which
the system has zero total energy and is given by $V_c^2=F(M_1,M_2,M_3)/d=
G M_1 M_2(M_1+M_2+M_3)/M_3(M_1+M_2)d$, where $M_1$ is the mass of the
primary, $M_2$ the mass of the secondary and $M_3$ is the mass of the
incoming, single star. For hard binaries, where $V_c \gg V_\infty$, 
the exchange cross section can be written as

\begin{figure}
\begin{center}
\begin{minipage}{125mm}
\plotfiddle{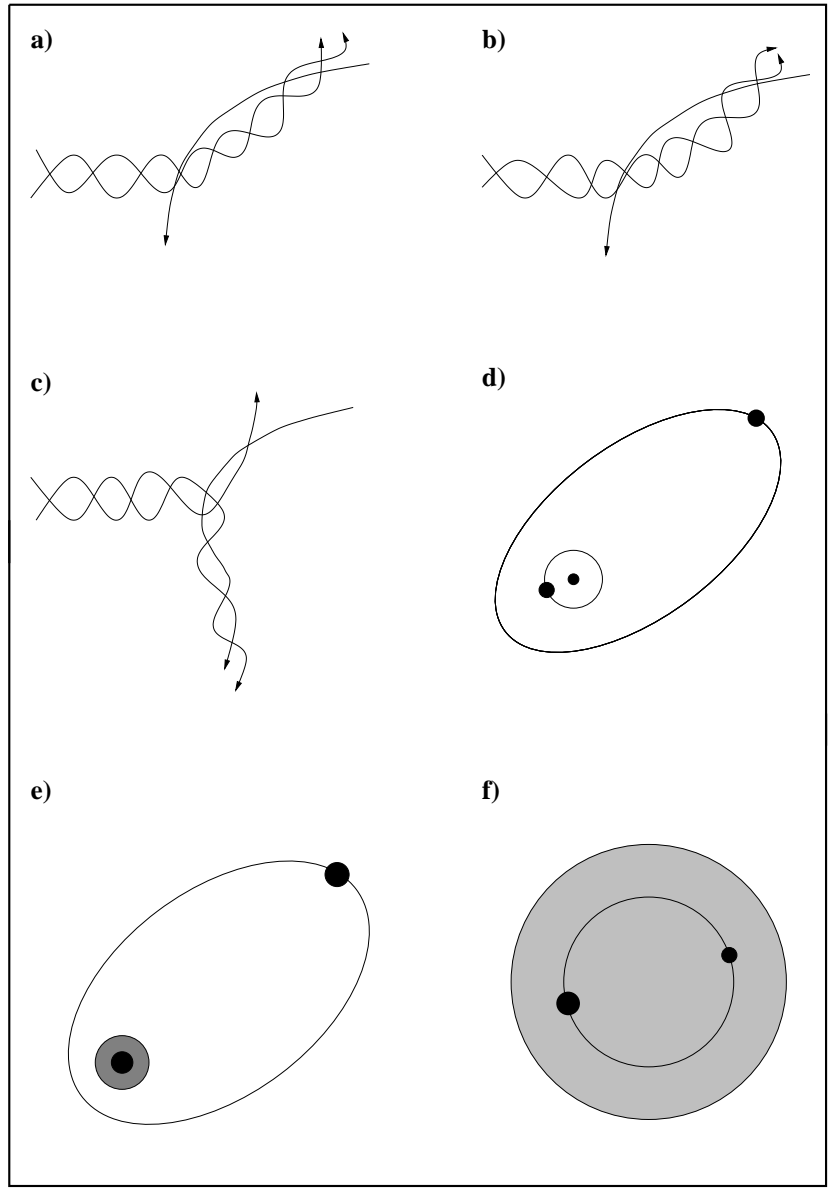}{16truecm}{0}{45}{45}{-155}{20}
\end{minipage}
\caption{\label{mdfig2} The possible outcomes of encounters
between binaries and single stars: a) a fly--by, b) a scattering--induced
merger where a fly--by leads to a merger of the two stars in the
binary, c) an exchange, d) a triple system, e) a merged binary where
two stars have merged and remains bound to the third star, and f)
a common envelope system where the envelope of the merged star engulfs
the third star.}
\end{center}
\end{figure}

\begin{equation}
\sigma_{\rm ex} = k_{\rm ex}(q_1,q_2) \pi d^2 
\cdot { G M_1 F(q_1,q_2) \over d} \cdot { 1 \over V_\infty^2 }
\end{equation}

\noindent where $q_1=M_2/M_1$, $q_2=M_3/M_1$, and
$F(q_1,q_2) = q_1(1+q_1+q_2)/q_2(1+q_1)$.  The constant $k_{\rm ex}(q_1,q_2)$
has to be determined through numerical simulations.

Similarly, one may write an expression for the cross section
for fly--by encounters, where the components of the binary remain unchanged,
but the binary is hardened by some amount.

\begin{equation}
\sigma_{\rm fb} = k_{\rm fb}(q_1,q_2) \pi d^2 \cdot { G M_1 F(q_1,q_2) \over
d} \cdot { 1 \over V_\infty^2 }
\end{equation}

\noindent where again $k_{\rm fb}(q_1,q_2)$ has to be determined through
numerical simulations.

One may also compute the cross section for two of the three stars to pass
within some minimum distance, $R_{\rm min}$, during an encounter.
Hut \& Inagaki (1985) found that such a cross section can be
written in the following form,

\begin{equation}
\sigma_{mb} = k_{\rm rmin}(q_1,q_2) \pi d^2 
{ G M_1 F(q_1,q_2) \over d } \cdot  { 1 \over V_\infty^2 }
\left( R_{\rm min} \over d \right)^\gamma
\end{equation}

\noindent where both $k_{\rm rmin}$ and $\gamma$ can be found through
simulations of encounters. For encounters involving three identical stars,
$\gamma \sim 0.5 - 1$.

The most likely outcome of an encounter between a single star
and a wide (though hard) binary is either a fly--by or an exchange.
Such an event will harden the binary by $\sim$ 20\%. After a number of
these encounters the binary will therefore be very much smaller
and consequently the relative probability of some variety of merger
occurring during later encounters will be enhanced. For example,
binaries containing solar--like stars which are
just resilient to breakup in a typical globular cluster
 have initial separations $d \sim 1000 R_\odot$
but will have separations $\sim 50 - 100 R_\odot$ today 
(see for example Davies \& Benz 1995).

In encounters between two binaries, we again require
the two systems to pass within $\sim d$ of each other. Hence 
binary/binary encounters will dominate over binary/single encounters
only if the binary fraction is $\geq 30\%$.
Unfortunately, the binary fraction in many stellar clusters is not well known.

\section{Importance of Collisions in producing Stellar Exotica}

Thus far we have seen that collisions and tidal encounters between
two single stars will occur in the cores of globular clusters 
and that encounters involving binaries will occur both in globular 
and open clusters. These encounters will be important for a number
of reasons. They may produce the various {\sl stellar exotica} that
have been seen in clusters such as
blue stragglers and millisecond pulsars. 
Stellar encounters will also have a role
in the dynamical evolution of clusters. Stellar collisions in clusters
may also lead to the production of massive black holes. Once produced,
these black holes may be fed by the gas ejected in subsequent stellar
collisions.

Given that each star involved in a collision may
be either a main--sequence star (MS), a red giant (RG), a white dwarf (WD), 
or a 
neutron star (NS), there are ten distinct combinations of collision
pairs as shown in 
Figure 3. In this review we will consider encounters
involving two main--sequence stars, and encounters involving 
a red giant or main--sequence star and a compact star ({\it i.e.\ } 
either a white dwarf or a neutron star). Encounters between two 
main--sequence stars may be responsible for the observed population
of blue stragglers within globular and open
clusters, as will be discussed in a later section.
Those involving compact objects and red giants or main--sequence
stars may produce interacting binaries where material is transferred
from the larger donor onto the compact object. Examples  of
interacting binaries include low--mass X--ray binaries (LMXBs) and 
cataclysmic variables (CVs) and both classes of objects will 
be considered in subsequent sections.

\begin{figure}[ht]
\vspace{-1.5truecm}
\begin{center}
\plotfiddle{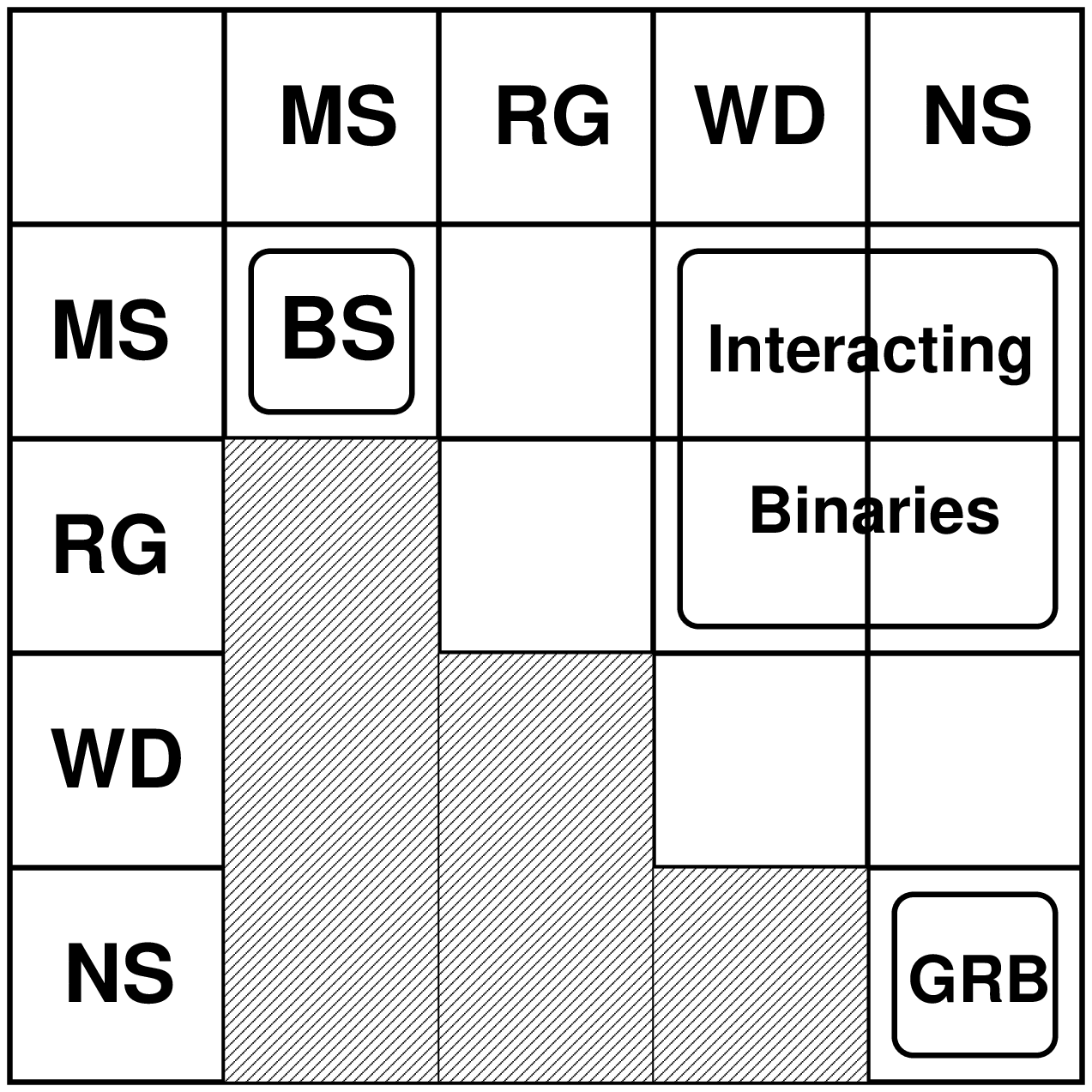}{7truecm}{0}{45}{45}{-100}{0}
\caption{\label{mdfig3} The various possible collisions between
two stars.}
\end{center}
\end{figure}

Merging neutron stars have been suggested as the source of 
gamma--ray bursts (GRBs). A direct collision between two neutron
stars is extremely unlikely. However, tight neutron--star binaries
might be produced through the  evolution of binaries,  including those
that have been involved in encounters with other stars. The neutron stars
in such tight binaries may then ultimately merge as they spiral together
as angular momentum and energy is emitted in the form of gravitational
radiation.  The event rate of such mergers may therefore be enhanced 
within dense clusters. We will not consider such mergers further in
this review.

\section{Blue Stragglers}

Blue stragglers are positioned on the upper end of the main--sequence
beyond the present day turn--off mass.
They have been observed in many globular clusters, including:
$\omega$~Cen (Kaluzny et al. 1997),
47 Tuc (Paresce et al. 1991), NGC 6397 (Auriere et al. 
1990),
M30 (Yanny et al. 1994, Guhathakurta et al. 1996),
 and
M80 (Ferraro et al. 1999).
These stars may have formed from the merger of two lower--mass
main--sequence stars either in an encounter between two single stars
or in encounters involving binaries
when two main--sequences collide and merge as part of the encounter.
The fraction of blue stragglers
in binaries may therefore be an important diagnostic for the binary fraction
for globular clusters.

Simulations of collisions between two main--sequence stars
have been performed by many groups, most using the method
known as Smoothed Particle Hydrodynamics (SPH).
SPH is a lagrangian method where the fluid
is modelled as an ensemble of particles  which follow the
flow of the fluid. Computational resources are not wasted in following
the evolution of the voids, such as the gaps between two colliding stars,
 and the resolution can vary in
a natural way; more particles being found in the places of most interest.
Because SPH has no specific need for a computational box, we are able to
follow the flow of gas completely. Thus we do not experience the
Columbus Effect,
where material is lost off the edge of a computational domain.

For a given value of the relative velocity, $V_\infty$, between
the two stars, one may compute how close the stars have to pass
in order for a capture to occur. Even closer encounters will
produce a single, merged object.
Simulations 
yield values for the capture radius, $R_{\rm capt} \sim 3R_{\rm ms}$,
and provide a lower limit for the merger radius 
of $R_{\rm merg} \sim 2R_{\rm ms}$.
The mass lost from the system on the initial impact is small,
typically $M_{\rm lost} \leq 0.01 M_{\rm ms}$.
Early work suggested that the merged stars would be well mixed
(Benz \& Hills 1987, 1992). Subsequent simulations, using a more
centrally concentrated model for the main--sequence stars, seem
to suggest that the material in the cores will not (at least initially)
mix with the envelope gas (Lombardi, Rasio \& Shapiro 1995).
More recently, the subsequent evolution of the merged objects 
has been considered and theoretical distributions on
a colour--magnitude diagram have been produced (Sills \& Bailyn
1999).  This study showed
that the distribution of blue stragglers in M3 
were difficult to reproduce using a single set of assumptions,
however if three particular bright blue stragglers were
neglected, the remaining population could have been produced
in mergers of stars occurring 
in encounters between binaries and single stars.
Extremely--high resolution SPH simulations of main--sequence star
collisions are ongoing to study in more detail the internal
structure of the merger products (see for example Sills et al. 2001).

\section{X--ray Binaries and Millisecond Pulsars}

Both millisecond pulsars (MSPs)
and low--mass X--ray binaries (LMXBs) have been observed in relative
abundance in globular clusters
clearly indicating that their origin is related to stellar encounters.
Under the standard model,
MSPs are produced in LMXBs where the neutron star
is spun--up as material is accreted from the Roche--filled companion. However
observations suggest that there are far more MSPs than LMXBs which,
given their comparable expected lifetimes, poses a problem for the
standard model. One would therefore wish to investigate whether 
encounters may lead to other potential channels for MSP production
which will not pass through a prolonged phase of X--ray emission.

Early work focussed on encounters between single neutron stars and either
red giants or main--sequence stars.
Fabian, Pringle \& Rees (1975) suggested that such encounters would
produce the observed X--ray binary population.
Calculations of encounter rates suggest that encounters involving
main--sequence stars will be more frequent than those involving 
red giants.
Numerical hydrodynamic simulations 
of encounters between neutron stars and red giants or main--sequence 
stars revealed that the $R_{\rm capt} \sim 3.5 R_\star$ for main--sequence
stars, and $R_{\rm capt} \sim 2.5 R_\star$ for red giants (Rasio \&
Shapiro 1991; Davies, Benz \& Hills 1992, 1993). Consideration
of the subsequent evolution of the two stars suggested a lower limit
of $R_{\rm merg} \ga 1.8 R_\star$ in both cases. As all
encounters in globular clusters are highly gravitationally focussed,
the cross section for two stars to pass within a distance $R_{\rm min}$,
$\sigma_{\rm rmin} \propto R_{\rm min}$. Hence approximately {\it half} of
the bound systems will form binaries and the rest will form single merged
objects. In the latter case encounters involving main--sequence
stars will produce a thick--disk system 
with the shredded remains of the main--sequence star engulfing the
neutron star. The equivalent encounters involving red giants will 
produce common envelope systems, where the red--giant envelope 
engulfs both the neutron star and the red--giant core.

Even if all the merged systems produced MSPs without passing
through a prolonged X--ray phase, the expected MSP production rate is 
only a factor of $\sim 2-3$ times larger than that for the LMXBs.
The solution to the MSP enigma seems unlikely to lie with encounters
involving single stars.

\begin{figure}
\begin{center}
\begin{minipage}{135mm}
\plotfiddle{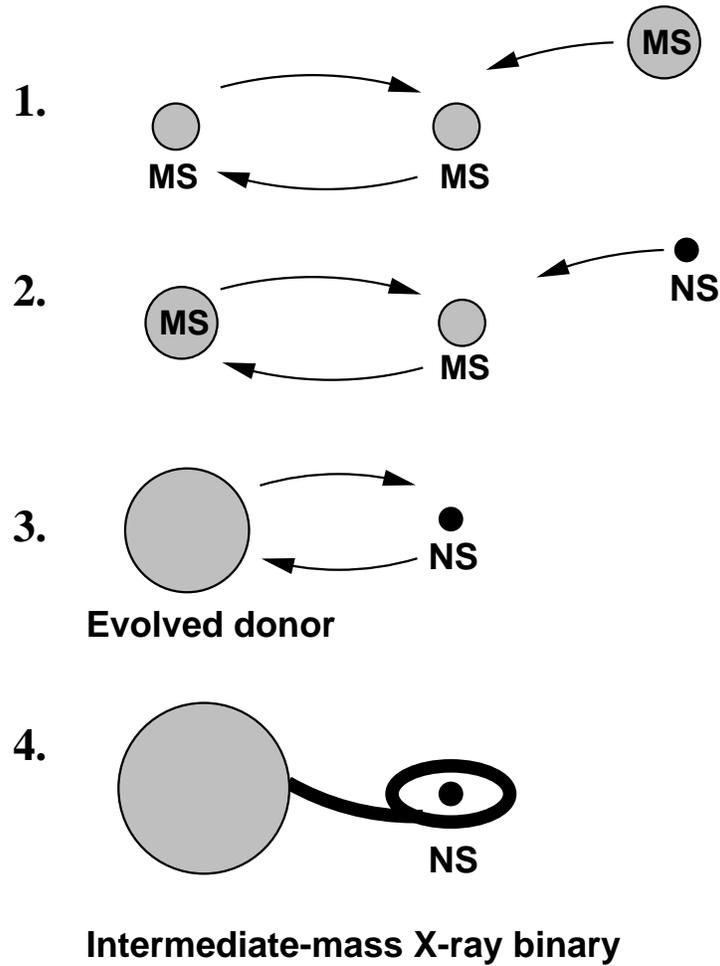}{14truecm}{0}{100}{100}{-290}{-220}
\caption{\label{mdfig5} The evolutionary pathway to produce intermediate--mass
X--ray binaries (IMXBs) in globular clusters (Davies \& Hansen 1998).
A more--massive main--sequence star exchanges into a binary containing
two main--sequence stars (phase 1), a neutron star exchanges into the
binary replacing the lower--mass main--sequence star (phase 2). The 
intermediate--mass star evolves of the main--sequence and fills its Roche
Lobe (phase 3). The system has a relatively short phase as an 
IMXB (phase 4), possibly producing a millisecond pulsar.}
\end{minipage}
\end{center}
\end{figure}


As was mentioned earlier, the binary fraction within globular clusters
is highly uncertain. However, because of the larger cross section for
encounters involving binaries, only a small fraction of binaries are
required for binary--single encounters to be as important as encounters
between two single stars. Calculation of the cross sections for 
fly--bys, exchanges, and mergers, lead to the predicted production 
rates for LMXBs and smothered neutron stars from encounters involving
primordial binaries (Davies, Benz \& Hills 1993, 1994; Davies \& Benz 
1995; Davies 1995). 
Although encounters today will produce smothered
neutron stars, they also produce LMXBs in roughly equal numbers;
the MSP production rate problem remains.

The solution may lie in considering the {\sl past}. 
The idea (developed by Davies \& Hansen [1998]) is shown in Figure 4.
Exchange encounters will
tend to leave the most massive stars within binaries, independent of 
the initial binary composition. When neutron stars exchange into
these binaries, the less massive of the two main--sequence stars
will virtually always be ejected. The remaining
main--sequence star will typically
have a mass of $\sim 1.5 -3 M_\odot$. The binary will evolve into
contact once the donor star evolves up the giant branch.

The subsequent evolution of such a system will depend on the mass
of the donor star and the separation of the two stars when the donor
fills its Roche lobe. For example, it has been suggested that the
system may enter a common envelope phase (eg Rasio, Pfahl \& Rappaport 2000).
Alternatively, the system may produce an {\sl intermediate--mass 
X--ray binary (IMXB)}.
In such a system the neutron star may accrete 
sufficient material (and with it, angular
momentum), for  it to acquire a rapid rotation (ie millisecond periods).
Because the donors
are all more massive than the present turn--off mass in globular
clusters, all IMXBs will have undergone their mass transfer
{\sl in the past}. If these systems evolve into MSPs, then we obtain,
quite naturally, what is observed today, namely a large MSP population
and a relatively small X--ray binary population.

Observations and modelling of the X--ray binary Cygnus X--2, 
provide important clues in helping determine the subsequent evolution of
intermediate--mass systems.
This binary is unusual in that its donor has the appearance (by its
position in an HR diagram) of a slightly--evolved 3--5 M$_\odot$ star,
yet its measured mass is much lower ($\sim 0.5$ M$_\odot$).
The evolutionary history of Cyg X--2 has been considered 
(King \& Ritter 1999, Podsiadlowski \& Rappaport 2000, and 
Kolb et al. 2000).
The unusual evolutionary state of the secondary
today appears to indicate that the system has passed through a period
of high--mass transfer from an initially
 relatively--massive star ($\sim 3.6 $M$_\odot$)
which had just evolved off the main sequence. The neutron star has somehow
managed to eject most of the $\sim 2 $M$_\odot$ of gas transferred
at Super--Eddington rates from the donor during this phase. This evolutionary
history  may also apply to the IMXBs formed dynamically in globular clusters.
Vindication of this model also comes from studying the dynamical evolution
of the binary within the Galactic potential (Kolb et al. 2000).
A suitable progenitor binary originating in the
Galactic Disc has sufficient time, and could have received
a sufficient kick when the primary exploded to produce a neutron star,
to reach the current position of Cyg X--2.

\section{Cataclysmic Variables}

A cataclysmic variable (CV) comprises of a low--mass star
in a binary with a white dwarf. The low--mass star is filling
its Roche lobe and is transferring material onto the white
dwarf via an accretion disc.
CVs are thus the white--dwarf analogues of LMXBs. One might therefore
imagine that the CV population within a globular cluster may be boosted
in a similar fashion to the LMXB population. 
However detecting CVs in globular clusters has proved to be difficult,
although a few clusters are now known to contain  
spectroscopically--confirmed CVs (for a review see Grindlay 1999).
Ongoing surveys of a number of globular clusters using Chandra and XMM
are also likely to boost the known population (see for example Cool et al.,
in these proceedings).

CVs might be produced via tidal encounters between white dwarfs and
main--sequence stars (Verbunt \& Meylan 1988).
Calculations of encounters
between binaries and single white dwarfs demonstrate that white dwarfs
will exchange into primordial binaries producing CVs (Davies 1995;
Davies \& Benz 1995). 
CVs are also produced in primordial binaries without the outside
interference of a passing single star. Indeed $\sim$ 1\% of all
binaries will produce CVs (de Kool 1992) 
by passing through the following
stages.
Beginning with two main--sequence stars in the binary,
the primary will evolve off the main--sequence, expand up the
giant branch and fill its Roche lobe. The subsequent mass transfer may
be unstable and lead to the formation of a common envelope of gas around
the red--giant core (which is essentially
 a white dwarf) and the secondary (which is still on the main sequence).
This enshrouding envelope of gas will be ejected as the white dwarf and
main--sequence star spiral together. If the initial separation of the
two stars were too small, the main--sequence star and white dwarf will
coalesce before all the envelope is removed, however under favourable
initial separations, the entire common envelope can be removed leaving
the white dwarf and main--sequence star in a tight binary (with a separation of
a few $R_\odot$). Such a binary will then come into contact if angular
momentum loss mechanisms can work on sufficiently short timescales or
when the main--sequence star  evolves into a red giant. A CV will then be
produced if the subsequent mass transfer is stable.

The formation route for {\sl primordial CVs (PCVs)}
described above will be {\sl inhibited} in dense
clusters if encounters with single stars or other binaries
disturb the PCV binary before the onset of the common--envelope
stage (Davies 1997),
for example an intruding third star 
may break up a wide binary. An interesting consequence is that PCVs are
unlikely to be found within the cores of dense clusters, but will
be found in their halos and throughout lower--density clusters. Conversely
CVs formed via encounters between binaries and single stars are likely
to be found exclusively in the cores (where the encounters will occur)
and will be produced in greater numbers within higher--density clusters
(Davies 1997) as  is illustrated 
schematically in Figure 5.

\begin{figure}[t]
\vspace{-3truecm}
\begin{center}
\plotfiddle{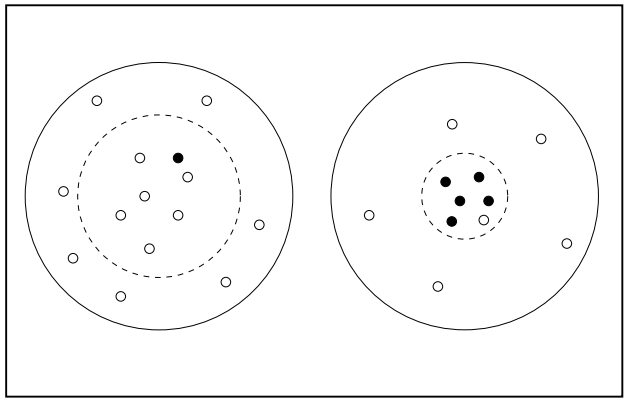}{8truecm}{0}{40}{40}{-115}{0}
\caption{\label{mdfig6} A schematic illustration of the CV population
within a low--density (left) and high--density (right) globular cluster. The 
dashed circles denote the cores of the clusters. Filled circles are CVs
formed dynamically, whilst open circles are primordial CVs (PCVs) 
which have formed from primordial binaries.}
\end{center}
\end{figure}

\section{Summary}

We have reviewed stellar encounters in crowded places, considering
their frequency and importance in producing the various exotic
objects seen in  clusters. Our conclusions
are as follows

\begin{enumerate}
\item Encounters happen in crowded places, such as the cores of
globular clusters, and galactic nuclei.
\item They may lead to the formation of various
observed {\sl stellar exotica}, such as blue stragglers.
\item Blue stragglers may be formed via mergers of lower--mass
main--sequence stars.
\item Interacting binaries such as LMXBs and CVs will be produced
via encounters between compact objects and either single stars or
binaries.
\item The MSP population in globular clusters may have been produced
from an earlier generation of {\sl intermediate--mass X--ray binaries (IMXBs)}.
\end{enumerate}

\acknowledgments
The support of the Royal Society through a University Research 
Fellowship is gratefully acknowledged.

\end{document}